\documentstyle[numreferences]{jobp}      

\def\square{\Box}

\begin{opening}
\title{A Clifford Dyadic Superfield from Bilateral Interactions
of Geometric Multispin Dirac Theory}
\author{William M. Pezzaglia Jr.}
\institute{Department of Physics\\
Santa Clara University\\
Santa Clara, CA 95053\\U.S.A.\\
Email: wpezzaglia@scuacc.scu.edu}
\author{Alfred W. Differ}
\institute{Department of Physics\\ American River College\\
Sacramento, CA 95841\\U.S.A.}
\date{(Received: Oct 1993)}
\end{opening}
\runningtitle{Preprint\# clf-alg/pezz9304}
\begin{document}

\begin{abstract}
Multivector quantum mechanics utilizes wavefunctions which are Clifford
aggregates (e.g. sum of scalar, vector, bivector).  This is equivalent to
multispinors constructed of Dirac matrices, with the representation independent
form of the generators geometrically interpreted as the basis vectors of
spacetime.  Multiple generations of particles appear as left ideals of the
algebra, coupled only by now-allowed right-side applied (dextral) operations.
A generalized bilateral (two-sided operation) coupling is proposed which
includes the above mentioned dextrad field, and the spin-gauge interaction
as particular cases.  This leads to a new principle of
{\it poly-dimensional covariance}, in which physical laws are invariant
under the reshuffling of coordinate geometry.  Such a multigeometric
superfield equation is proposed, which is sourced by a bilateral current.
In order to express the superfield in representation and coordinate free form,
we introduce Eddington E-F {\it double-frame} numbers.
Symmetric tensors can now be
represented as 4D ``dyads", which actually are elements of a global 8D
Clifford algebra.  As a restricted example, the dyadic field created by the
Greider-Ross multivector current (of a Dirac electron) describes both
electromagnetic and Morris-Greider gravitational interactions.

\keywords{} spin-gauge,  multivector,  clifford, dyadic
\end{abstract}

\section{Introduction}
Multivector physics is a grand scheme in which we attempt to describe all
basic physical structure and phenomena by a single geometrically interpretable
Algebra.  A conservative approach recognizes the Dirac algebra as belonging to
a
Clifford manifold having both spin and coordinate aspects.  The
{\it spin gauge theory} approach to grand unification makes use of a
spin Clifford algebra which necessarily commutes with coordinate geometry.
We propose a direct projection from this abstract space into concrete
coordinate geometric algebra.  Ultimately we eliminate spin space entirely
by using Clifford aggregates of coordinate geometry to replace
`spinors'.  Spin gauge theory, an artifact of spin geometry therefore vanishes.
However, we gain in having multiple generations of particles appear which are
coupled by new dextrad (right-sided multiplication) gauge transformations.

To accommodate all the known couplings we must somehow recover the spin-gauge
formalism.  This requires transformations which literally reshuffle the
geometry, i.e. the basis vectors for one observer might be the trivectors for
another observer.  This leads us to propose that the general physical laws are
invariant under these transformations, a new principle called
{\it poly-dimensional covariance}.  We postulate a single multigeometric
superfield equation, which will require two commuting coordinate Clifford
algebras, analogous to Eddington's E-F
``double frame" numbers\cite{Salingaros}.  This {\it dyadic}
Clifford algebra can be reinterpreted as a single 8D multigeometric space.
Multivector Dirac theory expressed in this full algebra potentially has
enough degrees of freedom to represent all the fermions of the standard model.

\section{Geometric Algebras and Multi-Spinors}
We present at first the standard view that abstract entities (e.g. spinors)
exist outside of the realm of concrete coordinate geometry.  Dirac algebra
belongs to a Clifford manifold which has both spin and coordinate features.
We propose a direct projection between spin space and coordinate geometry
in eq. (2) below.

\subsection{Spacetime and the Majorana Algebra}
Factoring the second-order {\it meta-harmonic} Klein-Gordon equation
to the first order {\it meta-monogenic} Dirac form requires four
mutually anticommuting
algebraic elements \(\{ \gamma^1,\gamma^2,\gamma^3,\gamma^4\}\),
$$   ({\Box}^2 - m^2)\Phi(x)=
({\bf \Box} - m)({\bf \square} + m)\Phi(x),  \eqno(1a)$$
$$ \Psi(x) = ( \Box + m)\Phi(x) =
(\gamma^{\mu}\nabla_{\mu}+m)\Phi(x), \eqno(1b)$$
$$ (\gamma^{\mu}\nabla_{\mu}-m)\Psi(x)=0, \eqno(1c)$$
where $\nabla_{\mu}=\partial_{\mu}$ in flat spacetime.
Requiring the formulation to be Lorentz covariant
imposes the defining condition of a Clifford algebra,
${1 \over 2} \{ \gamma_{\mu},\gamma_{\nu} \}=
g_{\mu \nu}= {\bf e}_{\mu} \bullet {\bf e}_{\nu}$,
where ${\bf e}_{\mu}$ are the coordinate basis vectors.  If the use of the
abstract $i$ is excluded, the above factorization of eq. (1a) only works
in the metric signature of $(+++-)$.  The lowest order matrix representation
of the $\{\gamma_{\mu}\}$ is ${\bf R}(4)$, i.e. 4 by 4 real
(i.e. no commuting $i$) matrices, commonly known as the (16 dimensional)
{\it Majorana algebra}.  The explicit matrix form of the algebra generator
$\gamma_{\mu}^{\ \alpha \beta}$ can be determined from the Riemann space
metric $ g_{\mu \nu}$ up to a similarity {\it spin transformation}:
$\gamma_{\mu}^{\prime}= S\gamma_{\mu}S^{-1}$.

\subsection{Spin Space}
The solution of the Dirac eq. (1c) is usually taken to be a four component
column {\it bispinor} $\Psi^{\alpha}$, belonging to the left linear space
for which the endomorphism algebra is the Majorana matrices.  This
{\it spin space} is transcendental, i.e. the postulates of quantum mechanics
ordain that some attributes (e.g. quantum phase) of the wavefunction cannot
be directly observed.  The {\it principle of representation invariance}
states that tangible results
should be invariant under a spin transformation:
$\Psi^{\alpha\prime} =S^{\alpha}_{\ \beta}\Psi^{\beta}$.
It should therefore be possible to express the theory in a form which
eliminates
any reference to a particular representation without sacrificing any
``physics".
To this end we introduce the {\it spinor basis}
${\bf \xi}_{\alpha}$ as carriers for the representation.
A spin transformation can now be interpreted as a passive
change in spinor basis, which leaves the {\it spin vector}
${\bf \Psi}=\Psi^{\alpha}{\bf \xi}_{\alpha}$ unchanged.

The {\it dual} spinor basis $\bar{\bf \xi}_{\alpha}$ is defined such that
$\bar{\bf \xi}_{\alpha} {\bf \xi}_{\beta}=\eta_{\alpha \beta}$,
where the spin metric $\eta_{\alpha \beta}$ has the diagonal signature $(++--)$
in the standard matrix representation.  We propose to interpret the
representation independent form,
$$ {\bf e}_{\mu} = {\bf \xi}_{\alpha}
\gamma_{\mu}^{\ \alpha \beta} \bar{\bf \xi}_{\beta},  \eqno(2a)$$
as the ``observable" basis vector of coordinate space.  Mathematically this
can be viewed as a map or projection from the Clifford manifold to the
coordinate manifold.
Hence we get a Dirac equation completely independent of spin basis or matrix
representation: \(({\boldmath \Box}-m){\bf \Psi}=0\), where
\({\bf \Psi} =\Psi^{\alpha} {\bf \xi}_{\alpha}\) and
\({\boldmath \Box}= {\bf e}^{\mu} \partial_{\mu}\)
is now the coordinate gradient.

\subsection{Geometric interpretations of Gauge Algebras}
There is a long standing tradition which views $i$ as only ``existing"
in spin space, as the internal ${\bf U}(1)$ generator of unobservable
quantum phase.  Factors of $i$ are included as needed to make
operators Hermitian
(e.g. $\gamma_4$) so that expectation values will never contain a
non-observable ``imaginary" number.  The usual Dirac matrices are the
complexified Majorana algebra: ${\bf C}(4)={\bf C} \otimes {\bf R}(4)$.
This can be geometrically reinterpreted as a 5D geometric (anti de-Sitter)
space, where the unit pseudoscalar (5-volume) plays the role of the
$i=\gamma^1\gamma^2\gamma^3\gamma^4\gamma^5$,
only if the fifth basis vector has positive signature.
The obvious question would be the physical interpretation of the new
fifth dimension, and the identification of its associated coordinate
variable and conjugate momenta (mass?).  We will address this question
briefly below.

To represent an isospin doublet of bispinors (e.g. u \& d quark) requires a
commuting isospin Pauli $\{\sigma_j \}$ algebra.  The wavefunction can
be expressed as a matrix set of components $\Psi^{\alpha \kappa}$
contracted on a product basis
${\bf \xi}_{\alpha} {\bf \lambda}_{\kappa}$.  There are only two elements to
the isospinor basis $\{ {\bf \lambda}_1, {\bf \lambda}_2\}$ which necessarily
commute with the spinor basis ${\bf \xi}_{\alpha}$.  The direct product of
a commuting Majorana ``spin" algebra and a Pauli ``isospin" algebra can be
reinterpreted as a 7D geometric algebra with metric signature $(+++----)$.
The column spinor for which the endomorphsim algebra is
${\bf C}(8)={\bf C}(2) \otimes {\bf R}(4)$ would now have 8 components.

To represent two observers in real spacetime requires a pair of coordinates,
each with their own Clifford algebras\cite{Differ}.
The direct product of these two commuting algebras can be geometrically
reinterpreted as a 8D space with a {\it mother algebra}\cite{Doran}
${\bf R}(16)={\bf R}(4) \otimes {\bf R}(4)$.
This encompases all the above algebras, where the `second frame'
${\bf R}(4)$ algebra commutes with that of the `first frame'.
Hence the `second' algebra is the
`internal' gauge algebra for the `first' frame observer and visa versa.

\section{Spin Covariant Dirac Theory}
The {\it special theory of relativity} requires the Dirac equation to have the
same form under Lorentz transformations: $dx^{\mu}=a^{\mu}_{\ \nu} dx^{\nu}$.
It is usually argued\cite{Drell} that the generators $\gamma^{\mu}$ are
invariant
scalars, i.e. the same for all observers, at the cost of forcing the bispinor
wavefunction to obey a compensating spin transformation:
$\psi^{\alpha \prime}=S^{\alpha}_{\ \beta} \psi^{\beta}$, where
$S^{-1}\gamma^{\mu}S= a^{\mu}_{\ \nu} \gamma^{\nu}$.

\subsection{Coordinate Covariant Dirac Theory}
The {\it general principle of covariance} will require the spin transformation
to be local (different at each point in spacetime).  This introduces a
{\it spin connection} ${\bf \Omega}_{\mu}$ to the derivative $\nabla_{\mu}$
of the Dirac eq. (1c),
$$ \nabla_{\mu}=\partial_{\mu} + {\bf \Omega}_{\mu},  \eqno(3a)$$
$$ \partial_{\mu} {\bf \xi}_{\alpha} = {\bf \Omega}_{\mu}{\bf \xi}_{\alpha} =
{\bf \xi}_{\beta}\Omega_{\mu \ \alpha}^{\ \beta}, \eqno(3b)$$
$${\bf \Omega}_{\mu}=\Omega_{\ \mu}^{(j)}\ {\bf E}_{(j)}=
\Omega_{\ \mu}^{(j)}\ {\bf \Gamma}^{\ \beta \alpha}_{(j)}
{\bf \xi}_{\beta} \bar{\bf \xi}_{\alpha}=\Omega_{\mu}^{\ \beta \alpha}
\ {\bf \xi}_{\beta} \bar{\bf \xi}_{\alpha}, \eqno(3c)$$
$${\bf \Omega}_{\mu}^{\prime}=
S\ {\bf \Omega}_{\mu} S^{-1} + S \partial_{\mu} S^{-1}. \eqno(3d)$$
One of the 16 basis elements ${\bf E}_{(j)}$ of the geometric Clifford
algebra is given by the generalization of eq. (2a),
$${\bf E}_{(j)}=\Gamma_{(j)}^{\ \ \alpha \beta}
\ {\bf \xi}_{\alpha} \bar{\bf \xi}_{\beta}, \eqno(2b)$$
where $\Gamma_{(j)}$ is the corresponding basis element of
the Dirac matrix algebra.
Under the general coordinate transformations required by the
{\it equivalence principle}, one must replace
$\gamma^{\mu} \rightarrow \gamma^a h^{\ \mu}_a(x)$ where the {\it tetrad}
(vierbein) field $h^{\ \mu}_a(x)$ transforms as a vector.  This is
equivalent to introducing position dependent $\gamma^{\mu}(x)$ which
transform like basis vectors.

For this reason and others, we adopt the ``nontraditional" view that both
\( {\bf e}^{\mu} \) and $\gamma^{\mu}$ of eq. (2a) transform as vectors, while
${\bf \xi}_{\beta}$ and $\bar{\bf \xi}_{\alpha}$ transform as coordinate
scalars\cite{Chapman}.  With this definition of constant
spin basis, the spin connection is everywhere zero, hence the
generally covariant Dirac equation is simply eq. (1c) with Dirac matrices
which are a function of position.  However, when the coordinate space is
curved,
one cannot have the spin connection vanish everywhere.  The geometric
definition of eq. (2a) forces the following relations:
$$\partial_{\nu} \gamma_{\mu} = C_{\nu \mu}^{\ \ \omega}\  \gamma_{\omega}-
{\bf \Omega}_{\ \nu}^{(j)} [{\bf \Gamma}_{(j)},\gamma_{\mu} ],  \eqno(4a)$$
$$ [ {\bf K}_{\omega \sigma},{\bf e}_{\mu} ] =
R_{\omega \sigma \mu}^{\ \ \ \ \nu} {\bf e}_{\nu}, \eqno(4b)$$
$$ {\bf K}_{\omega \sigma}= K_{\omega \sigma}^{(j)}\  {\bf E}_{(j)} =
[\nabla_{\omega},\nabla_{\sigma}], \eqno(4c)$$
The coordinate connection coefficient $C_{\nu \mu}^{\ \ \omega}$
(Christoffel symbol) is directly related to the spin connection by eq. (4a).
Restricting our discussion to real spacetime algebra (no commuting $i$),
the {\it spin curvature} \({\bf K}_{\sigma \omega}\) is forced by eq. (4b) to
be a bivector.  Clearly it must be nonzero if the coordinate space is curved,
i.e. described by the Riemann curvature tensor:
$R_{\omega \sigma \mu}^{\ \ \ \ \nu}{\bf e}_{\nu}=
[\partial_{\omega},\partial_{\sigma}]{\bf e}_{\mu}$.
It follows from eq. (4c)
that the spin connection (which appears in the spin covariant derivative
\(\nabla_{\sigma}\)) must have a nontrivial bivector part, commonly called
the {\it Fock-Ivanenko coefficient}\cite{Chapman}.

\subsection{Spin Gauge Theory}
The principle of {\it local matrix representation invariance} or equivalently
a principle of {\it spin basis covariance} is invoked to induce via minimal
coupling a non-trivial spin connection\cite{Enjalran}.  This is a gauge theory
where the generators \( {\bf \Gamma}_{(j)}\) of the general spin transformation
are usually restricted to be Dirac bar-negative in order to preserve the spin
norm $\bar{\Psi}\Psi$ (i.e. the spin metric
$\bar{\bf \xi}_{\alpha} {\bf \xi}_{\beta} = \eta_{\alpha \beta}$
is invariant).  The standard (5D) Dirac algebra which has the bar negative
pseudoscalar $i$, would contain the 16 element group structure $U(2,2)$.
Electromagnetism is associated with $i$, which by itself would force the
space curvature to be zero.  It is tempting to interpret the 10 bivectors
(of 5D) with group structure $SO(4,1)$ as the gauge fields which cause
gravitational curvature through eq. (4b).

Grand unification is approached by Chisholm and Farwell\cite{Chisholm}
by resorting to higher dimensions (e.g. 11D) to introduce more fields.
They only consider spin transformations of the form:
$\gamma^{\mu}=\gamma^a h^{\ \mu}_a(x)$, generated by bivectors or the
pseudoscalar $i$.  They avoid those bivectors which would rotate
spacetime into a higher dimension (e.g. $\gamma^5\gamma^1$).  The remaining
bivectors which operate on spacetime form the 6 element Lorentz group
$SL(2,C)$, potentially insufficient to accommodate a full description
of gravitation.

\subsection{Local Automorphism Invariance}
Alternatively, the entire automorphism group $U(2,2)$ of the Dirac algebras is
allowed by Crawford\cite{Crawford}.  Previously, the non-bivector generators
were excluded by equations (4a) \& (4b).  These constraints are relaxed because
Crawford does not require the geometric interpretation of eq. (2ab).  This
allows him to consider generalized spin transformations of the form:
\( {\bf \Gamma}^{(j)}= {\bf \Gamma}^a \Delta^{\ (j)}_a(x) \), where
$ {\bf \Gamma}^{(j)}$ is a basis element
of the full ``spin" (Dirac) Clifford algebra.  The
{\it drehbein} fields \({\bf \Delta}_a^{\ (j)}\)(x) (``spin-legs") reshuffle
multivector rank in the Clifford spin manifold (e.g. vector $\leftrightarrow$
bivector) without doing the same to the ``observable" coordinate geometry.

A Lagrangian formulation can show that the field equation is,
$${\bf K}_{\mu \nu}^{\ \ \ ; \mu} + [{\bf \Omega}^{\mu},{\bf K}_{\mu \nu}]=
{\bf j}_{\mu} = j_{\ \mu}^{(i)}\  {\bf \Gamma}_{(i)}, \eqno(5a)$$
$$j_{\ \mu}^{(i)}= {1 \over 2}\bar{\Psi}\{{\bf \Gamma}^{(i)},
\gamma_{\mu} \} \Psi ={1 \over 8} Tr(\Psi \bar{\Psi} \{{\bf \Gamma}^{(i)},
\gamma_{\mu} \}), \eqno(5b)$$
The current \({\bf j}_{\mu} \) is similar to the spin gauge connection
\( {\bf \Omega}_{\mu} \) in being a coordinate vector
while also a Clifford aggregate
over the spin algebra \({\bf \Gamma}^{(i)} \).  The spin curvature
{\bf K} can be geometrically interpreted as a dyad of a {\it coordinate}
geometric bivector and a {\it spin algebra} Clifford aggregate.
$${\bf K}=K^{\mu \nu}_{\ (i)}\ {\bf e}_{\mu}\wedge{\bf e}_{\mu}
\ {\bf \Gamma}^{(i)}. \eqno(5c)$$
Elements of the coordinate geometry \({\bf E}_{(i)} \) commute with the spin
algebra \({\bf \Gamma}^{(i)} \) because Crawford does not postulate the
geometric connection of eq. (2ab).  Note that the bivector part of the spin
curvature is no longer constrained by eq. (4b) to be related
to the space curvature.

\section{Multivector Gauge Theory}
The basic difference from standard theory is the replacement of column
spinors by algebraic wavefunctions, i.e. Clifford aggregates of Dirac
matrices\cite{Greider}.  Most authors only consider restricted
combinations called {\it minimal ideals},
which have the same degrees of freedom as a single column spinor.  In our
approach, the form of the multivector wavefunction is unrestricted, having the
same number of degrees of freedom as the elements of the Clifford group.  The
complete solution can be interpreted as a geometric multispinor:
\({\bf \Psi} = \Psi^{(i)} {\bf E}_{(i)} =
\Psi^{\alpha \beta} {\bf \xi}_{\alpha}  {\bf \lambda}_{\beta} \).  Here the
${\bf \xi}_{\alpha}$ is no longer a basis spinor, but an element of a left
ideal, hence eq. (3b) is no longer valid.  The isospin element is part of the
same algebra:  ${\bf \lambda}_{\beta}=\bar{\bf \xi}_{\beta}$,
which does not commute with ${\bf \xi}_{\alpha}$ whereas it did in standard
formulation.  In  4D spacetime algebra (no commuting $i$) the geometric
multispinor has been shown\cite{Pezz92} to be an isospin doublet of Dirac
bispinors, where the role of $i$ is played by right-side applied
(i.e. {\it dextrad multiplication}) time basis vector ${\bf e}_4$.
In 5D (standard Dirac algebra) one has enough degrees of freedom to
represent four quarks (i.e. u,d,s,c), where the (u,d) and (s,c) isospin
doublets are uncoupled.

\subsection{Dextral Gauge Theory}
The generally covariant multivector Dirac equation
$({\bf e}^{\mu} \partial_{\mu}- m)\Psi^{(i)} {\bf E}_{(i)}=0$, where
${\bf e}_{\mu}(x)$ are the local coordinate basis vectors, is manifestly
matrix representation independent.  We have in fact completely eliminated spin
space, specifically spin basis ${\bf \xi}_{\alpha}$ and spin algebra
$\gamma_{\mu}^{\ \alpha\beta}$ in favor of there being only
the geometrically interpretable coordinate Clifford algebra ${\bf E}_{(i)}$.
Hence, spin gauge theory, an artifact of spin space, is now inaccessible!

The multiple particle generations in the multivector wavefunction
can be coupled by now-allowed right-side applied
{\it dextral gauge transformation}\cite{Pezz9303}.  The new gauge fields enter
as a {\it dextrad connection}: \({\bf D}_{\mu}=D_{\mu}^{\ (i)}{\bf E}_{(i)}\),
$$ \nabla_{\mu} (\Psi) = \partial_{\mu} \Psi + \Psi {\bf D}_{\mu}, \eqno(6a)$$
coupling to the multivector parts of Greider's
current\cite{Greider},
$$ j_{\ \mu}^{(i)} = Tr({\bf E}^{(i)} \bar{\Psi} {\bf e}_{\mu} \Psi ) =
Tr(\Psi {\bf E}^{(i)}  \bar{\Psi} {\bf e}_{\mu} ). \eqno(6b)$$
A Lagrangian formulation\cite{Pezz92} will require the geometric generators
\({\bf E}_{(i)}\) of the dextrad connection\({\bf D}_{\mu}\) to be bar
negative.
In 4D spacetime, the subset which is also unitary generates the electroweak
group: $U(1) \otimes SU(2)$, where isospin rotations are generated by
spacelike bivectors and the role of $i$ played by right-sided (dextrad)
multiplication of the time basis element \({\bf e}_4\).

\subsection{Poly-Dimensional Covariance}
The spin gauge formalism can be recovered by proposing that the automorphism
transformations operate on the very real, concretely observable spacetime
coordinate Clifford algebra:
\({\bf E}_{(i)}^{\prime}={\bf E}_{(j)} \Delta^{(j)}_{\ \ (i)}(x)\).
The {\it geobein} fields $\Delta(x)$ (``geometry-legs") are completely
analogous
to Crawford's {\it drehbeins}\cite{Crawford} except that now we are reshuffling
observable geometry.  We are tautologically committed to propose a new
principle
of {\it local poly-dimensional covariance}.  By this we mean that the basis
vectors of a coordinate frame displaced from  the origin may be ``rotated" in
dimension, e.g. be a multivector that is part vector plus part bivector
relative
to the reference geometry.

The generalized {\it poly-dimensional connection} $\Lambda_{(i)}^{\ \ (j)}$
is defined,
$$ \Box {\bf E}_{(i)} = \Lambda_{(i)}^{\ \ (j)} {\bf E}_{(j)}. \eqno(7a)$$
The right side of this equation is recognized as a linear transformation on
the full Clifford algebra \({\bf R}(4)\).  In general $\Lambda_{(i)}^{\ \ (j)}$
belongs to the endomorphism algebra
$End \ {\bf R}(4) \cong {\bf R}(4) \otimes {\bf R}(4)$, hence it is an element
of the {\it mother} algebra ${\bf R}(16)$ \cite{Doran}.  This leads to a new
generalized {\it poly-dimensional covariant Dirac equation},
$$[(\Box - m)\Psi^{(j)} + \Psi^{(i)}
\Lambda_{(i)}^{\ \ (j)} ]{\bf E}_{(j)}=0, \eqno(7b)$$
where the coordinate gradient in eq. (7b) is understood now NOT to operate on
\( {\bf E}_{(j)}\).
This is not a particularly useful form, as it is expressed in terms of the
{\it multivector} basis \({\bf E}_{(j)}\) instead of an {\it ideal} basis which
would more closely resemble standard spinor form.  The main annoying
feature is that each multivector piece of the wavefunction couples to a
different connection coefficient.  Further, the poly-dimensional connection
cannot itself be expressed as a multivector within
the ${\bf R}(4)$ spacetime algebra.

Alternatively, the linear transformation can be written entirely within
the smaller original ${\bf R}(4)$ spacetime algebra using two-sided
multiplication\cite{Sommen}.  We re-express eq. (7a) in terms of a
new {\it bilateral connection} $\Omega^{(jk)}$,
$$ \Box {\bf E}_{(i)} = \Omega^{(jk)}
{\bf E}_{(j)} {\bf E}_{(i)} {\bf E}_{(k)}. \eqno(8a)$$
The advantage of eq. (8a) over eq. (7a) is that the connection is now
completely
form independent of the operand element ${\bf E}_{(i)}$.
This allows us to rewrite the interaction term of the Dirac equation in terms
of the full multivector wavefunction ${\bf \Psi}$ instead of having to consider
each multivector component $\psi^{(j)}$ separately as was done in eq. (7b).
The resulting Dirac equation has the
{\it bilateral interaction term} which was proposed earlier to empirically
fit known mesonic couplings\cite{Pezz9302},
$$ (\Box -m) {\bf \Psi}= -{\bf E}_{(i)}
{\bf \Psi} {\bf E}_{(j)} \Omega^{(ij)}, \eqno(8b)$$
where again it is understood that the gradient does not operate on the
multivector basis (as that has already been included on the right side of the
equation).  From a multivector Lagrangian formulation\cite{Pezz9302} it can
be shown that the gauge connection $\Omega^{(ij)}$ of eq. (8a) couples to
the {\it bilateral current},
$$ j^{(ij)} =
{1 \over 4} Tr( \bar{\bf \Psi} {\bf E}^{(i)}{\bf \Psi} {\bf E}^{(j)} )=
{1 \over 4} Tr( {\bf \Psi}
{\bf E}^{(j)}\bar{\bf \Psi} {\bf E}^{(i)} ), \eqno(8c)$$
where ${\bf E}^{(i)}$ and $ {\bf E}^{(j)}$ must both be bar-positive or both
bar-negative.  The dextral interactions of eq. (6a) are the special case where
the {\it sinistrad}\cite{Pezz9302} (left-side applied) interaction element of
eq. (8b) is the set of basis vectors: ${\bf E}_{(i)}={\bf e}_{\mu}$.  When the
{\it dextrad} (right-side applied) element ${\bf E}_{(j)}$ of eq. (8b)
is either $1$ or $i$,  the interactions are of the same form proposed by
Crawford\cite{Crawford}.

\subsection{Multivector Field Theory}
In order to have a fully geometric description of symmetric tensors,
Greider\cite{Morris} introduced a second commuting Clifford algebra
\({\bf F}_{(k)}\) in analogy with Eddington's E-F {\it double-frame}
numbers\cite{Salingaros}.  A product of two
elements \({\bf E}_{(j)}{\bf F}_{(k)}\) is a {\it geometric dyad} which is an
element of a global 8D {\it mother}\cite{Doran} algebra:
${\bf R}(16)={\bf R}(4) \otimes {\bf R}(4)$.
Potentially this allows us to write a single
superfield equation which is completely  coordinate and poly-dimensional
covariant in form.  In the particular case of dextrad connection
of eq. (6a), the superfield equation can be written in a
sourced {\it monogenic} form,
$$\Box {\cal F} = {\cal J}, \eqno(9a)$$
where ${\cal J}=j^{\mu (j)}{\bf f}_{\mu}{\bf E}_{(j)}$
is the vector-multivector supercurrent made from eq. (6b).  The coordinate
derivative is in the ${\bf F}_{(j)}$ algebra vector basis:
$\Box={\bf f}^{\mu}\partial_{\mu}$.  The superfield
${\cal F}=F^{\mu \nu (j)} {\bf f}_{\mu}\wedge {\bf f}_{\nu}{\bf E}_{(j)}$
is a bivector in the ``first-frame {\it coordinate} algebra" ${\bf F}_{(k)}$,
while a Clifford aggregate in the ``second-frame {\it charge}
algebra" ${\bf E}_{(j)}$.

The Morris-Greider\cite{Morris} theory of gravitation was based upon the
particular case where ${\bf E}_{(j)}$ is limited to be a vector,
the supercurrent then being a vector-vector dyad.
The case where ${\bf E}_{(j)}$ is a trivector was explored by
Differ\cite{Differ}.   It appears that the spin-gauge field
eq. (5a) can also be written in this general form, where the commutator term
is built into the equation if assumptions are made about the generalized
connection coefficient of eq. (7a).  The field equation for the general
bilateral interaction of eq. (8b) has yet to be fully formulated.

\section{Summary}
Our development was based upon an underlying theme of using only algebra that
is based on concrete spacetime geometry.  This has led us to eliminate spin
space,
the principle of local spin covariance and ultimately spin gauge theory.  In
its
place we propose the more grand scheme of
{\it local poly-dimensional covariance}.
While the results are promising for Dirac, gauge and classical field
theory, it remains to be seen if its domain can be extended to classical
mechanics.  Further, the interpretation of the 8D geometry needed is not
completely clear, although it appears to be connected with the classical
symmetric tensor objects of 4D which are needed for formulations of
gravitation.

\acknowledgements
This work was supported in part by the creative environment fostered
by the staff and proprietor of the
{\it Owl and Monkey Cafe} of San Francisco.  Many animated discussions
within contributed to this paper's completion, n.b. poet Larry Guinchard
(suggestion of name {\it poly-dimensional covariance}), Craig Harrison
(SFSU Philosophy Dept.) regarding conceptual foundations, and graduate student
John Adams (SFSU Physics Dept.) for providing a sounding board.
Appreciation also to M. Stein (Kaiserslautern, Germany) for advice on
appropriate generalizations of german terminology ({\it geobein}).
Finally, thanks to James Crawford (Penn State Fayette Physics Dept.) for
continued patience in arguing about the foundations of spin-gauge theory.


\begin{thebibliography}{99}

\bibitem {Chisholm} Chisholm, J.R.S. and Farwell, R.:1989, `Unified Spin Gauge
Theory of Electroweak and Gravitational Interactions',
{\it J. Phys.} {\bf A22}, 1059-71.

\bibitem{Crawford} Crawford, J.P.:1993, `Local Automorphism Invariance:
A Generalization of General Relativity', {to appear in}
{\it Proceedings of the Third International Conference on Clifford
Algebras and Their Applications in Mathematical Physics} {R. Delanghe,
F. Brackx and H. Serras, eds., Kluwer Academic Publ. 1993).}

\bibitem{Pezz9303} Pezzaglia, W.M.:1993, `Classification of Multivector
Theories
and the Modification of the Postulates of Physics', {to appear in}
{\it Proceedings of the Third International Conference on Clifford
Algebras and Their Applications in Mathematical Physics}, {R. Delanghe,
F. Brackx and H. Serras, eds., Kluwer Academic Publ. 1993).}

\bibitem{Enjalran} Enjalran, M.:1993, `Spin Gauge theory of Pauli Particles',
{Master's thesis, San Francisco State University.}

\bibitem{Differ} Differ, A.W.:1991, `The Gyric Theory: A Classical Field Theory
for Spin', {Doctorial thesis, University of California at Davis.}

\bibitem{Greider} Greider, K.:1984, `A Unifying Clifford Algebra Formalism for
Relativistic Fields', {\it Found. Phys.} {\bf 14}, 467-506;
:1980, `Relativistic Quantum Theory with Correct Conservation Laws',
{\it Phys. Rev. Lett.} {\bf 44}, 1718-21; `Erratum', {\it Phys. Rev. Lett.}
{\bf 45}, 673.

\bibitem{Salingaros} Salingaros, N.:1985, `Some Remarks on the Algebra
of Eddington's E Numbers', {\it Found. Phys.} {\bf 15}, 683-691.

\bibitem{Morris} Morris, F.G. and Greider, K.R.:1986, `The Importance of
Meaningful Conservation Equations in Relativistic Quantum Mechanics for
the Sources of Classical Fields', {\it Proceedings of the NATO and SERC
Workshop
on Clifford Algebras and Their Applications in Mathematical Physics},
Chisholm, J.S.R. and Common, A.K. eds.,
{\it NATO ASI Series} {\bf C183}, 455-63.

\bibitem{Chapman} Chapman, T.C. and Leiter, D.J.:1976, `On the generally
covariant Dirac Equation', {\it Amer. Jour. Phys.} {\bf 44}, 858-62.

\bibitem{Drell} Bjorken, J.D. and Drell, S.D.:1964, {\it Relativistic Quantum
Mechanics}, {McGraw-Hill, New York}.

\bibitem{Pezz92} Pezzaglia, W.:1992, `Clifford Algebra Geometric-Multispinor
Particles and Multivector-Current Gauge Fields',
{\it Found. Phys. Lett.} {\bf 5}, 57-62.

\bibitem{Pezz9302} Pezzaglia, W.:1993,
`Dextral and Bilateral Multivector Gauge Field description of
Light-Unflavored Mesonic Interactions', {Preprint: CLF-ALG/PEZZ9302}.

\bibitem{Sommen} Sommen, F. and Van Acker, N.:1993,
`SO(m)-Invariant Differential Operators on Clifford-Algebra-Valued
Functions', {(Preprint: CLF-ALG/SOMM9202), To appear in }
{\it Foundations of Physics}.

\bibitem{Doran} Doran, C., Hestenes, D., Sommen, F., and Van Acker, N.:1993,
`Lie groups and spin groups', {\it J. Math. Phys.} {\bf 34} 3642-69.

\end{thebibliography}
\end{document}